\documentclass[a4paper, 10pt, conference]{ieeeconf}      % Use this line for a4                                                          % paper
\IEEEoverridecommandlockouts                              % This command is only
\usepackage{graphicx}
\usepackage{float}
\usepackage{xcolor}
\title{\LARGE \bf
SmartEAR: Smartwatch-based Unsupervised Learning for Multi-modal Signal Analysis in Opportunistic Sensing Framework}
\author{Debanjan Borthakur$^{1}$,  Andrew Peltier$^{1}$, Harishchandra Dubey$^{2}$, Joshua Gyllinsky$^{1}$ and Kunal Mankodiya$^{1}$% <-this % stops a space$
\thanks{\textcolor{blue}{This material is presented to ensure timely dissemination of scholarly and technical work. Copyright and all rights therein are retained by the authors or by the respective copyright holders. The original citation of this paper is: D. Borthakur,  A. Peltier, H. Dubey, J. Gyllinsky and K. Mankodiya, "SmartEAR: Smartwatch-based Unsupervised Learning for Multi-modal Signal Analysis in Opportunistic Sensing Framework",  The Third IEEE/ACM Conference on Connected Health: Applications, Systems and Engineering Technologies, IEEE/ACM CHASE 2018, Sept. 26-28, 2018, Washington, D.C., USA. }}
\thanks{$^{1}$Wearable Biosensing Lab, University of Rhode Island, Kingston, RI, USA
{\tt\small  kunalm@uri.edu}}
\thanks{$^{2}$Center for Robust Speech Systems, The University of Texas at Dallas, Richardson, TX, USA.
}
}
\begin{document}
\maketitle
\thispagestyle{empty}
\pagestyle{empty}
\begin{abstract}
Wrist-bands such as smartwatches have become an unobtrusive interface for collecting physiological and contextual data from users. Smartwatches are being used for smart healthcare, telecare, and wellness monitoring. In this paper, we used data collected from the $AnEAR$ framework leveraging smartwatches to gather and store physiological data from patients in naturalistic settings. This data included temperature, galvanic skin response (GSR), acceleration, and heart rate (HR). In particular, we focused on HR and acceleration, as these two modalities are often correlated. Since the data was unlabeled we relied on unsupervised learning for multi-modal signal analysis. We propose using \textit{k}-means clustering, GMM clustering, and Self-Organizing maps based on Neural Networks for group the multi-modal data into homogeneous clusters. This strategy helped in discovering latent structures in our data. 
\end{abstract}
\vspace{0.2cm}
\section{INTRODUCTION}
\vspace{0.2cm}
Smartwatches are multi-purpose computerized wristwatches. They can collect information from internal and external sensors which include heart rate, blood pressure, oxygen saturation, acceleration, galvanic skin response, and more, depending on the model. Because they support wireless technologies like Bluetooth, Wi-Fi and GPS and can connect to smartphones to exchange sensor data, they can be used effectively for many functions~\cite{dubey2016bigear,barik2018geofog4health}. They are useful in tracking health and physical activity, and getting notifications extended from connected smartphones. As such, smartwatches play an important role in many healthcare situations. {\AA}rsand and colleagues discuss the use of smartwatches as a diabetes patient self-management tool \cite{aarsand2015performance}.
% ^.
Systems involving two-way communication between smartwatches and the mobile phones have promising possibilities for monitoring blood glucose and physical activity~\cite{priyadarshini2018investigation}. Other exciting and promising smartwatch applications include EchoWear, as presented in \cite{dubey2015echowear,dubey2015multi,mahler2016use,dubey2017fog}. EchoWear is a smartwatch system used for voice and speech treatments for patients with Parkinson's disease.
This article aims at the analysis of Multi-modal smartwatch signals that can be used for neurological assessment of the subject wearing the smartwatch. Opportunistic Sensing (OS) is a paradigm for signal and information processing in which a network of sensing systems automatically discover and select sensor platforms based on an operational scenario by determining an appropriate set of features and optimal means for data collection based on these features~\cite{liang2014opportunistic}. $SmartEAR$ is a use case of opportunistic sensing using smartwatches assisted with gateway devices such as smartphones/tablet and the secured cloud backend.

In this paper we are using a smartwatch-based data collection framework called $AnEAR$ \cite{constant2016anear}. It gathers physiological information from patients and stores it. This data can later be retrieved for unsupervised analysis, the main theme of this paper.
%There are several types of unsupervised clustering techniques exist.
%We have used \textit{k}-means, Gaussian Mixture Models and self-organizing maps to see the applicability of the $AnEAR$ framework. 
% What does this mean? 
Physiological signals collected from $AnEAR$ have been analyzed applying unsupervised grouping techniques, including \textit{k}-means, Gaussian Mixture Models and self-organizing maps.

%  \vspace{-1cm}
\begin{figure*}[t]
\centering
\includegraphics[width=450pt]{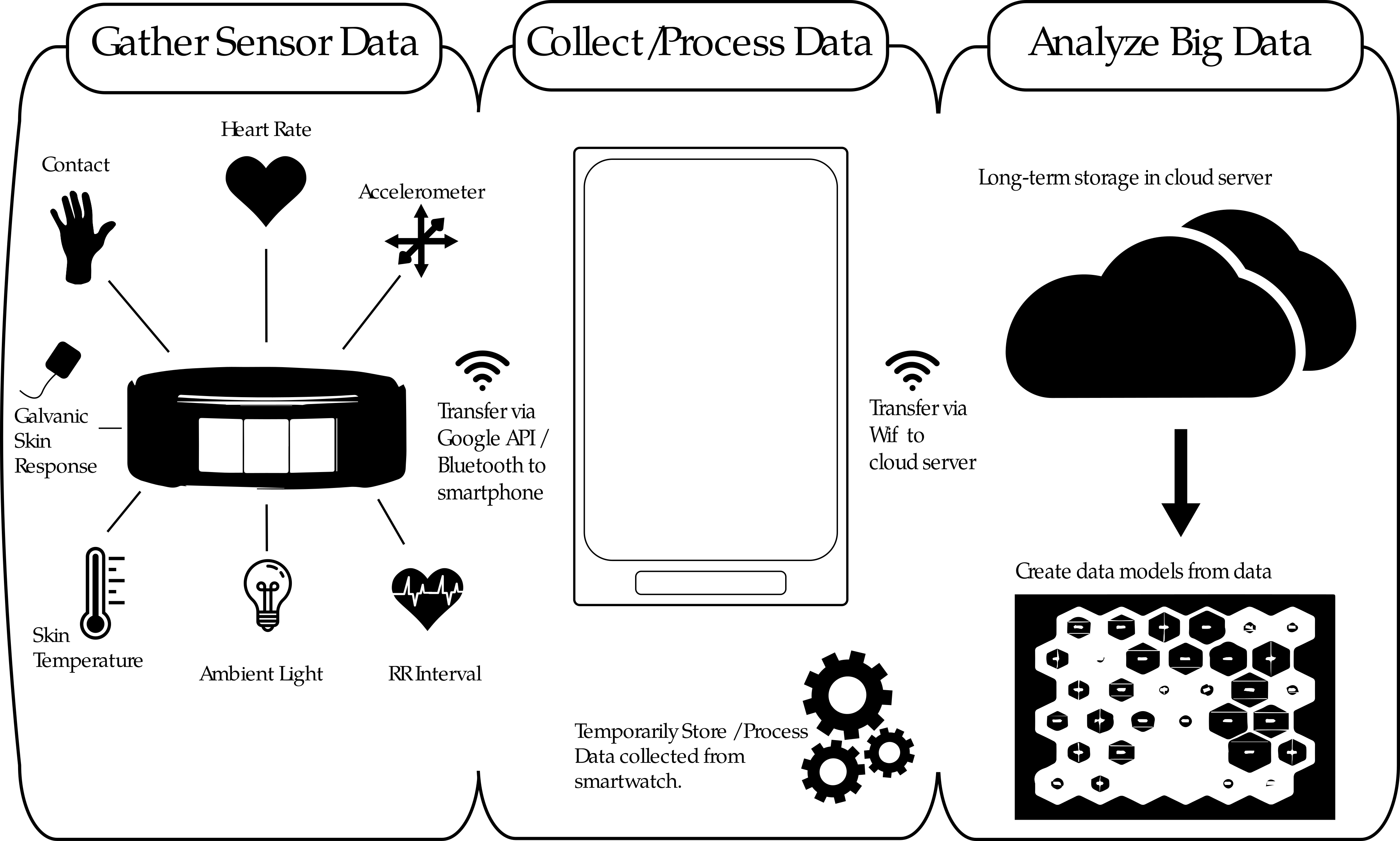}
%\includegraphics[width=175mm, height=60mm]{updatedFigure1.png}
% \vspace{-4mm}
\caption{Figure shows the $AnEAR$ framework with the smartwatch, android mobile and the unsupervised learning as application.}
\label{fig1}
\end{figure*}
%\vspace{-.5cm}
% \vspace{1.1cm}
\vspace{0.2cm}
\section{Related work}
\vspace{0.2cm}
\subsection{$AnEAR$}
$AnEAR$ (Android Electronically Activated Recorder) is a smartwatch-based data collection framework used to gather and store data from patients in their home environment. 
The framework consists of an Android smartphone, a smartwatch, and a server where the data is stored and analyzed using the $SmartEAR$ analysis framework presented in this paper.
Along with collecting physical data, $AnEAR$ uses its smartphone's audio recorder to record the patient's surrounding audio when activated by a heart rate based trigger.
In our current studies, patients use the Microsoft Band as the smartwatch component, as the Microsoft Band collects more accurate and plentiful than other smartwatches.
Smartwatch's sensors collect data including HR, GSR, acceleration, skin temperature, and ambient light levels. 
%Data includes acceleration, sensors include an accelerometer. 
At present, the $SmartEAR$ clustering analysis uses only HR and accelerometer data.

$AnEAR$ is currently being used in  studies to analyze data from patients with PTSD and other anxiety disorders.
Currently, $AnEAR$ attempts to identify panic attacks by triggering audio recordings when the patient's heart rate exceeds a predetermined threshold.
However, every patient is different, each having different resting heart rates, exercise habits, and triggers, creating individual variability in the data.
For this reason, the $AnEAR$ framework must be able to determine when each specific user is experiencing a spike in heart rate that is not attributed to exercise or movement using Machine learning and Deep learning. This has proven to be a challenge. Because patients are not in observable environments while participating in this study, it is impossible to tell what the patient is doing during data collection, and the data is therefore unlabeled.
Due to the impracticality of gathering labeled data, unsupervised algorithms are the only reasonable option for meaningful data analysis.
%Panic attacks cannot reliably be labeled as such when we are uncertain whether or not the patient was actually experiencing one.
%What is the purpose of that sentence?
For this reason, improvements to the framework's anxiety-detecting capabilities rely on unsupervised clustering analysis in order to identify any possible patterns that we can use to develop an effective Shallow Machine and Deep learning algorithm.
\vspace{0.2cm}
\subsection{Wearables and IoT}
\vspace{0.2cm}
\begin{figure}[h!]
\centering
\includegraphics[width=230pt]{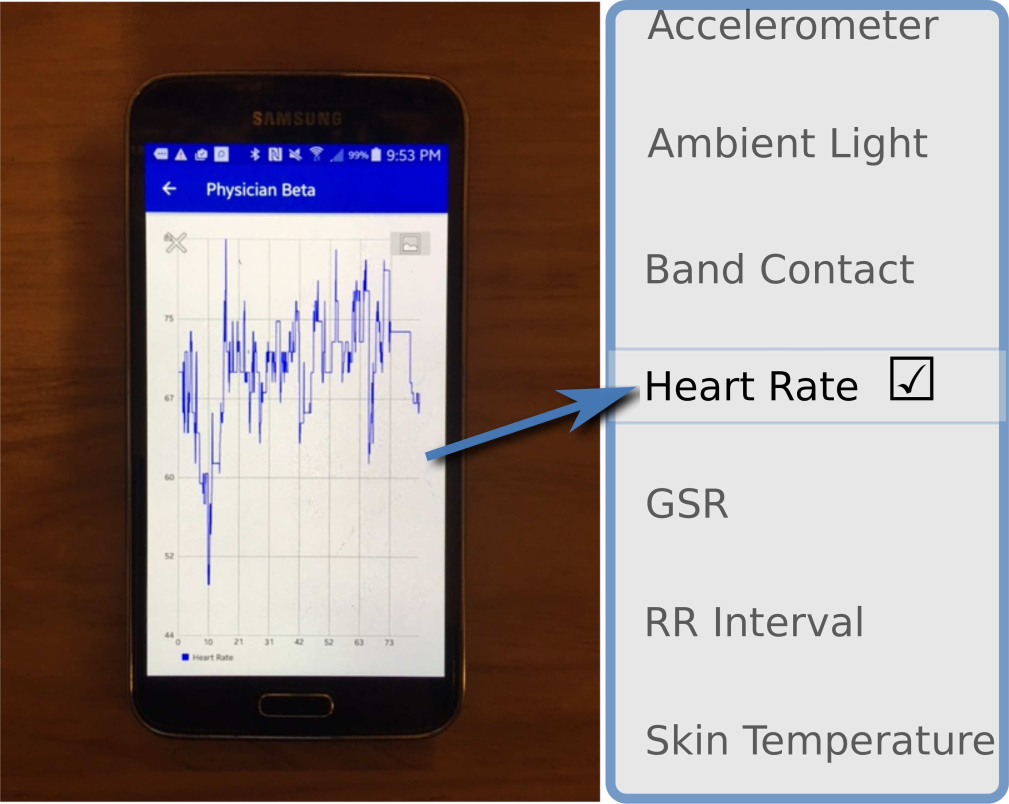}
\caption{AnEAR as a remote viewer (proper authentication required).}
\label{fig:AnEAR_ExpertView}
\end{figure}

Wearable technology and IoT device is an interconnection of sensors and fog-nodes that make them capable of transmitting and collecting the data \cite{samson2015wearing,constant2017fog,barik2017fog2fog}.
The big data from IoT devices can be used effectively for health monitoring.
The effectiveness of telehealth monitoring is seen in devices like EchoWear \cite{dubey2015echowear}.
Parkinson patients can use smartwatch in their speech exercises for monitoring purpose. Various authors proposed several architecture for IoT and Fog.
Kapur and colleagues describe the use of wearable sensor technology in real-time musical signal processing.
The system modifies resulting sounds based upon the movements of the performing artist \cite{kapur2005wearable} .
Emotion recognition through the use of physiological signals from the autonomic nervous system collected using wearables is another application \cite{lisetti2004using}. Features were mapped to emotions such as sadness and anger.
%\cite{lisetti2004using} authors aim to recognize users emotion by collecting the physiological signals from the autonomic nervous system using wearables and mapped them to certain emotions like sadness,anger etc.
WIoT architectures may also be presented in terms of design, function, and application \cite{r11_hiremath2014wearable}, such as in the FIT architecture \cite{r6_monteiro2016fit,barik2017soa,barik2018fog}.
%In \cite{r6_monteiro2016fit} authors proposes an FIT architecture where as in \cite{r11_hiremath2014wearable} authors present WIoT in terms of design,function and application.
It should, however, be noted that IoT applications require low latency and may encounter network bandwidth issues.
%Requirement of low latency, issues of network bandwidth are some of them. 

In our work, we leveraged the IoT based $AnEAR$ framework and proposed a model for predictive analysis using multi-modal smartwatch data.
The smartwatch component of $AnEAR$ is used to gather data, which it sends to the smartphone component via a Bluetooth Low-Energy connection. 
After the data is collected and processed, the smartphone sends its data to a cloud server for storage and further analysis.
This analysis includes our Deep learning and Shallow Machine learning algorithms. In the future this will allow us to create patient-specific models to identify panic attacks and other neurological events.
This model can be sent back to the smartphone over a secure network, which can be used to locally identify panic attacks. 
The biggest concern with the IoT structure as it relates to this or any other health study is data security, especially during data transfer.
Because patient data is sensitive by nature, securing that data is of utmost importance.
If the network that the patient's medical records are being transferred through is compromised, the records are at risk of being stolen.
Currently, to comply with HIPAA standards, we are manually transferring patient data via a secure thumb drive from the location of the study to our area of work in order to analyze the data.
\vspace{.2cm}
%\vspace{-1pt}
%\vspace{1cm}
%
\subsection{Multi-modal Smartwatch data}
\vspace{.2cm}
AnEar uses Microsoft Band to collect data from the subjects.
The Microsoft SDK exposes data from the sensors as streams, and applications can subscribe to this sensor streams. 
The table below shows some of the available sensor streams their sampling rates.

%\vspace{-.5cm}
\begin{table}[!h]
\centering
\caption{Description of various modalities in the smartwatch data.}
\begin{tabular}{|c|c|}
\hline
\textbf{Modality} & \textbf{Sampling Rate} \\\hline
 Heart rate (HR)  & 1 Hz \\ \hline% %
 Accelerometer & 8 Hz \\ \hline% %
 %Temperature & Core body temperature & \degree F \\ \hline% %
 Galvanic skin response (GSR) & 5 Hz  \\ \hline% %
 Ambient Light & 2 Hz \\ \hline% %
\end{tabular}
\label{table_modality}
\end{table}
%\vspace{-1cm}

Heart rate frequently correlates with acceleration data. 
For example, running and otherwise exercising subjects experience increases in both HR and acceleration. The figure below shows a correlation plot between 3 dimensional acceleration and heart rate data collected from '$AnEAR$'. A higher Positive correlation is seen with Pearson's rank correlation coefficient with the subject Heart rate and acceleration in X and Y direction although the value is small as we are not aware of the activities or tasks which makes this a completely unsupervised problem. This information can be vital in accessing the health status of the individuals. 
%Subjects in a running state will have both heart and accelerometer data going up at the same time which will result higher correlation between these two parameter.
%Subject 3 and subject 10? Write it out, since you haven't explained this abbreviation.
\vspace{2mm}
\begin{figure}[!h]
\centering
\includegraphics[width=260pt]{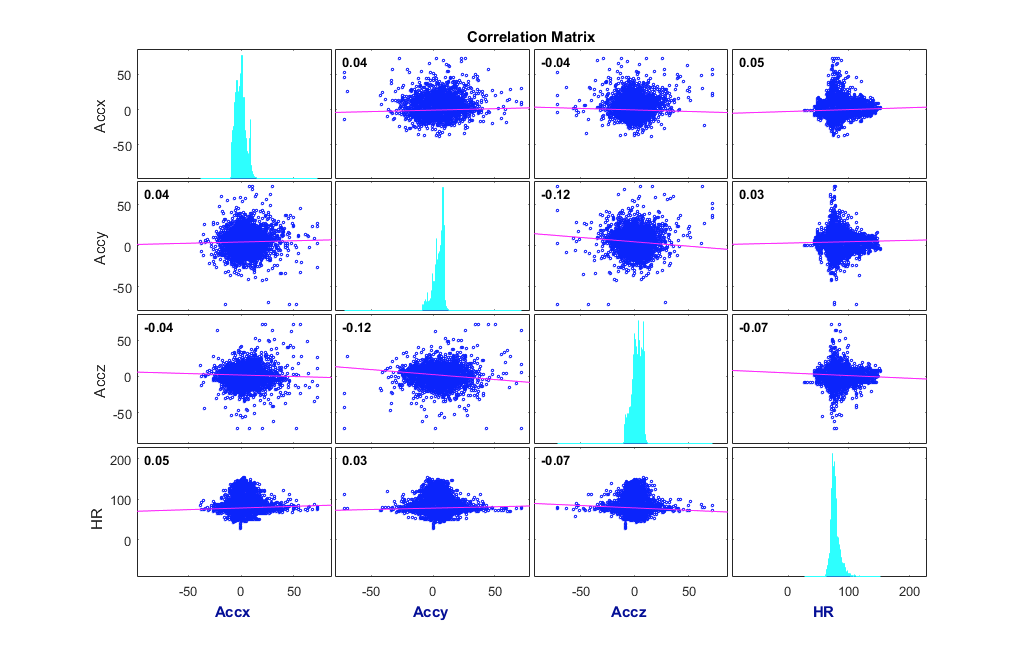}
%\vspace{-4mm}
\caption{Correlations between pairs of variables for one subject. Histograms are shown along the matrix diagonal; scatter plots of variable pairs appear off the diagonal. The displayed correlation coefficients are the slopes of the least-squares reference lines in the scatter plots.}
\label{fig2}
\end{figure} % This figure is not legible at the current size. 
\vspace{0.2cm}
\subsection{Data Collection}
\vspace{0.2cm}
Data was collected from each subject over a two week period. During the two week study period, participants wore a Microsoft Band. Data was recorded in three minute blocks, with three minutes without recording between each block. This data was sent to a paired Android device periodically throughout the day. Data from 10 participants is analyzed in this paper.

% * <alyssazisk@uri.edu> 2018-06-08T15:07:47.477Z:
%
% > Data was gathered from a subject, over the course of two weeks. 
% > During those two weeks, the patient would wear the Microsoft Band and send data to its paired Android device periodically throughout the day every day.
% > The data would be repeatedly recorded for three minutes, followed by a three minute resting period.
% > We have analyzed 10 subjects data in this paper. The correlation plot shows comparative analysis of two subjects individual data.
%
% ^.
\section{Proposed Approach}

\subsection{Unsupervised Learning}

Hastie, Tibshirani, and Friedman explain unsupervised learning as `learning without a teacher' \cite{hastie2009unsupervised}.    
Unsupervised machine learning is defined as the machine learning task of inferring a function to describe hidden structure from unlabeled data.
As the data is unlabeled, the accuracy of structures the algorithm reveals is not evaluated.
%there is no evaluation of the accuracy of the structure that is the output of the algorithm.

\begin{figure}[!h]
\centering
\includegraphics[width= 250pt]{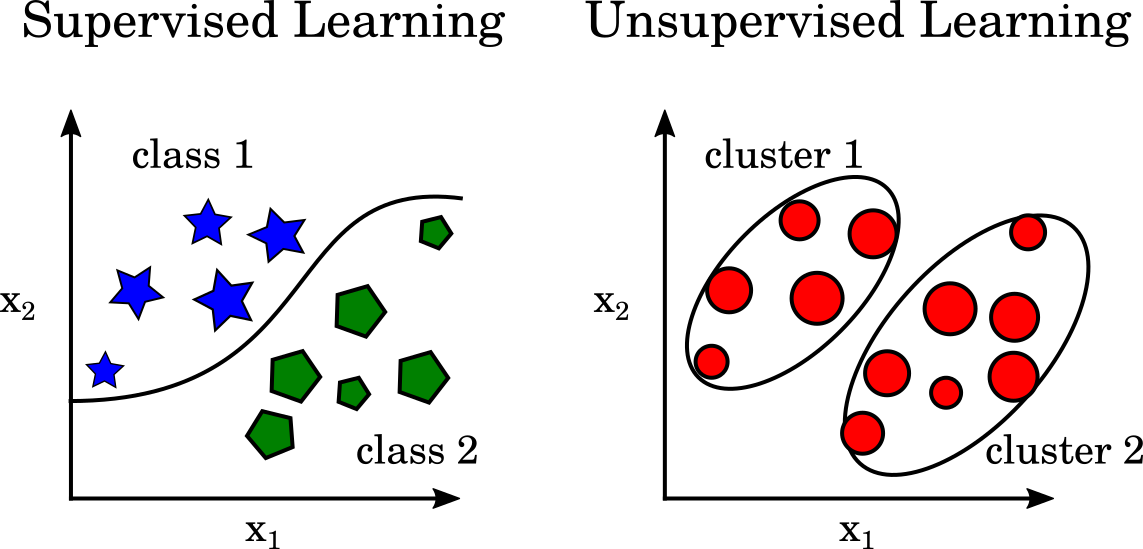}
%\vspace{-2mm}
\caption{The difference between supervised and unsupervised learning adapted from \cite{George} is shown.}
\label{fig3}
\end{figure}

Unsupervised learning tries to draw inferences from the dataset.
Cluster analysis is the most common form of unsupervised learning. It is often used for exploratory data analysis to find hidden patterns.
Clusters are usually modeled using a measure of similarity, such as the Euclidean or probabilistic distance.
Clustering algorithms include:

\begin{itemize}
\item Hierarchical clustering
\item \textit{K}-means clustering
\item Gaussian Mixture Models
\item Self-organizing maps
\item Hidden Markov models
\end{itemize}

We have used $AnEAR$ to collect and store unlabeled data and applied \textit{K}-means, Gaussian Mixture Models and Self-organizing maps for unsupervised clustering of the data.
\vspace{0.2cm}
\subsection{\textit{K}-means clustering}
\vspace{0.2cm}
\textit{K}-means clustering is an unsupervised learning that is used for exploratory data analysis of unlabeled data.
It is a method of vector quantization which is quite extensively used in data mining.
The algorithm finds $k$ groups in the data. In this method, observations are placed in the cluster with the nearest mean. This mean typically serves as a prototype for the cluster.
The \textit{k}-means++ algorithm uses an heuristic to find centroid seeds for \textit{k}-means clustering.
%\textit{K}-means clustering aims to partition observations into different clusters in which each observation belongs to the cluster with the nearest mean usually which serves as a prototype of the cluster.
The data space is thus partitioned into Voronoi cells.
As mentioned in \cite{borthakur2017smart} this algorithm aims to minimize the squared error function J given by:
\[
J=\sum^K_{k=1} \sum_{i\in c_k} || x_i - m_k ||^2 \ \ \ \ \ \ \ \rm 
\]
The Euclidean distance, squared Euclidean distance, Mahalanobis distance, or Cosine distance between data points and cluster centers can be used as the distance measure, which is then minimized.
Matlab functions are used to generate \textit{k}-means plots of the data sourced from the AnEar framework.
% The \textit{k}-means function uses the squared Euclidean distance metric and the \textit{k}-means++ algorithm for cluster center initialization. - redundant? I think you already said this.
%
\vspace{0.2cm}
\subsection{Gaussian Mixture Model clustering}
%\vspace{0.2cm}
%
Other approaches for clustering involve the use of certain models for clusters and attempt to optimize the fit between the data and the model.
A mathematical distribution, such as a Gaussian or Poisson distribution, can represent each cluster. The entire data set is then modeled by a mixture of these distributions.
GMM clustering is a fuzzy or soft clustering method, and is therefore inherently flexible.
This model is often useful when analyzing subpopulations within a given population. It breaks down a general data observation of a cluster into multiple, more specific observations.
This is incredibly useful when identifying the nuances of an anxiety disorder.
With lots of training data, we can potentially design the SmartEAR analysis to pick up subtle changes in mood that indicate a potential panic attack in a user by applying this mixture model to our analysis. 
% Panic attacks aren't all that subtle, but warning signs before they happen can be. Do you mean something more like this:
% With lots of training data, we can potentially apply this mixture model to the SmartEAR data to pick up subtle changes in mood that may be warning signs for user panic attacks.
%
\vspace{0.2cm}
\subsection{Self Organizing maps}
Self Organizing Maps (SOMs) are a form of a neural network that maps inputted values on a two-dimensional plane to create similar clusters of data. Priento and colleagues  proposed a new method for the detection and recognition of traffic signs using a SOM \cite{prieto2009using}. Their method first detects potential road signs by analysing the distribution of red pixels within images, and then it identifies the actual road signs from the distribution of dark pixels.
The SOM map is trained to sort and organize inputs into clusters. All values in the same cluster will then be processed and handled in the same way through cluster analysis.
In order to successfully organize the data in this manner, SOMs require a large amount of data to process. This allows its learning algorithm to be trained to identify and organize similar inputs, much like how the human brain processes sensory information by organizing different parts of the human brain to process different senses.
According to the European Journal of Operational Research \cite{mingoti2006comparing}, SOM networks make fewer observations than other analysis algorithms, including \textit{k}-means, when processing non-overlapping data sets, and outliers in the processed data don't seem to disturb our algorithms as much. Unfortunately, the rate of correct classification drops significantly for SOMs as the number of clusters increases.
%Since our SOM has created three clusters, while our \textit{k}-means algorithm has only created two, it is possible that the observations created by our SOM are less detailed. 
% The figures show 3 clusters for k-means and 2 clusters for GMM, neither of which is SOM. Cut the sentence above, rewrite it to apply correctly to SOM, or rewrite it to match what you did + place it in a section applicable to  k-means and GMM instead of SOM.
%
%\vspace{0.2cm}
\section{Results and Discussions}
%\vspace{0.2cm}
%
We collected data from the $AnEAR$ framework, including GSR, Heart Rate, Accelerometer data, and Temperature. This work presents the use of $AnEAR$ for clinical analysis. 
Multi-modal unlabeled smartwatch data can't be used to directly predict the user's activity or the neurological state.  However, there may be correlations between physiological signals acquired via $AnEAR$ which, will also aid understanding of possible relationships between different vital signals. We performed unsupervised clustering analysis in this study to show a use case for the $AnEAR$ smartwatch system. Three different techniques of unsupervised clustering were performed on the data collected from the $AnEAR$ system. We have chosen to focus on two variables, HR and acceleration, due to their greater likelihood of correlation.
\begin{figure}[!h]
\centering
\includegraphics[width=260pt]{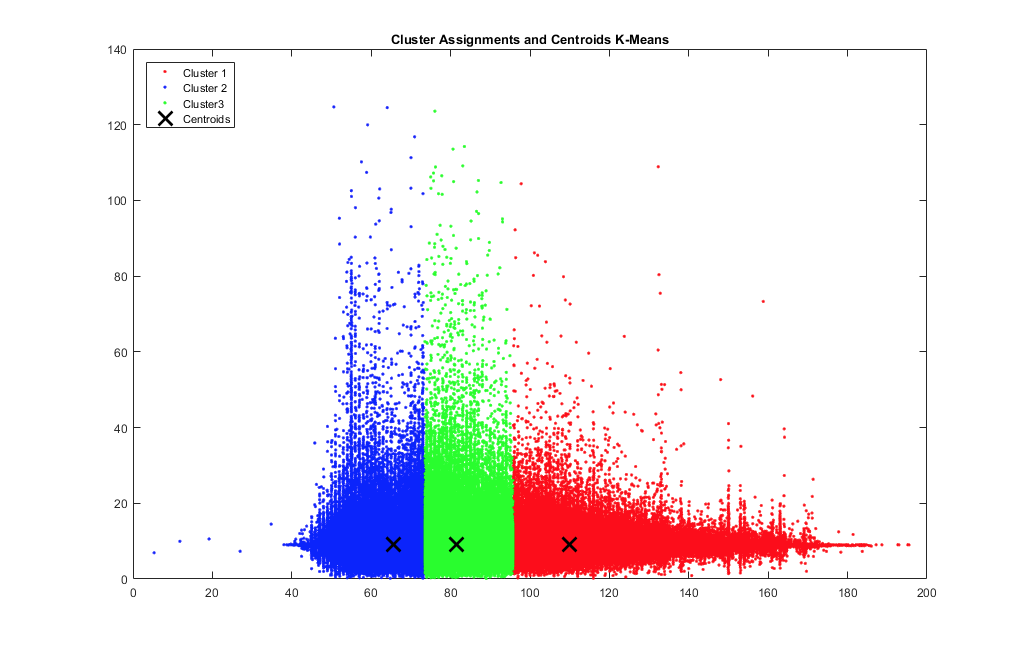}
%\vspace{-4mm}
\caption{\textit{K}-means clustering analysis is performed on $AnEAR$ data from 10 subjects.}
\label{fig4}
\end{figure}

Figure \ref{fig4} shows our implementation of \textit{k}-means clustering using the \textit{k}-means++ algorithm for centroid initialization and squared Euclidean distance. A cross marks centroid locations.
The data have been partitioned into three clusters using the best arrangement out of several initializations. Matlab \textit{k}-means by default initializes the replicates separately using \textit{k}-means++. It reduces the probability of a bad initialization leading to a bad clustering result. 
Initial cluster centroid positions were chosen to perform a preliminary clustering phase on a random 10 percent subsample of the input data using the Matlab command 'cluster.'
Here the value of K is three, so we see three clusters with different colors along with their centroids. We have used the same data and analyzed it using Gaussian Mixture Model clustering.

\begin{figure}[!h]
\centering
\includegraphics[width=260pt]{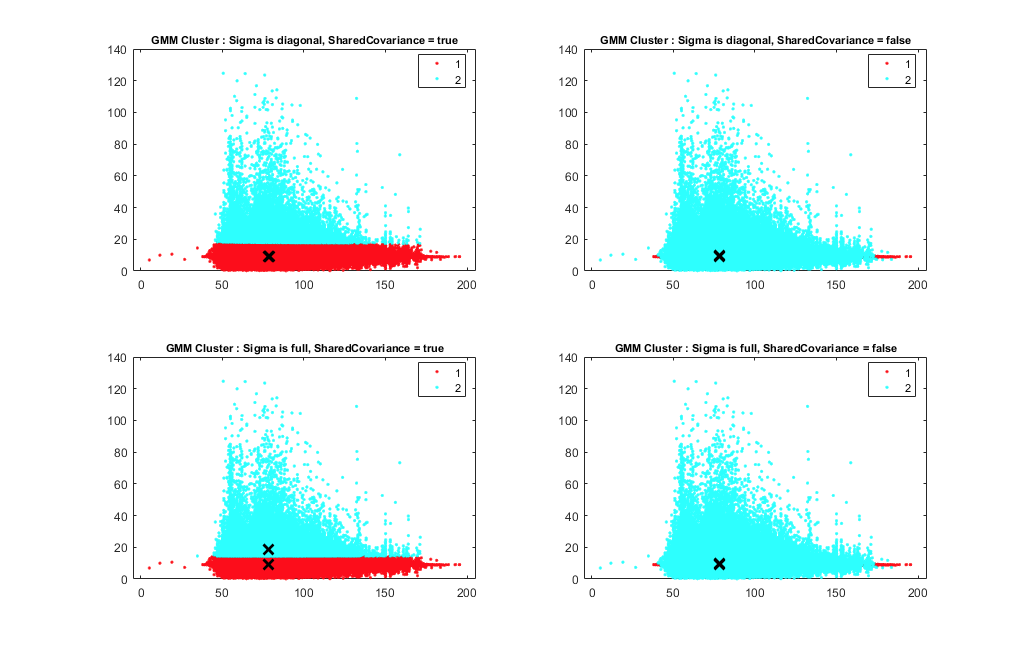}
%\vspace{-4mm}
\caption{GMM clustering analysis performed on $AnEAR$ data}
\label{fig5}
\end{figure}
%This figure isn't really legible at the current size either.
GMM clustering works by maximizing the component posterior probability given the data.
% GMM clustering is flexible as we can view it as a fuzzy or soft clustering method. 
%You already said this in the GMM section.
Figure \ref{fig5} shows the four plots generated by GMM clustering.
\begin{itemize}
\item Diagonal covariance matrices indicate that the predictors are uncorrelated.  
\item Full covariance matrices allow for correlated predictors. 
\item Shared covariance matrices mean that all components have the same covariance matrix.  
\item Unshared covariance matrices mean that all parts have their own covariance matrix.
\end{itemize} 
Unlike the \textit{k}-means, 2 Clusters are shown for GMM conducted on the $AnEAR$ data, as designated by two different colors.
The clusters are more or less similar and informative as shown in \textit{k}-means in Figure~\ref{fig4}. The only difference is the number of clusters used in the unsupervised learning. Another exploratory analysis we have used in the self-organizing maps using Neural Network.
% * <jgyllinsky@my.uri.edu> 2018-06-08T00:01:41.489Z:
% 
% This is a repeat
% > Another exploratory analysis we have used in the self-organizing maps using Neural Network.
% > \begin{figure}[h!]
% > \includegraphics[width=90mm]{image6.png}
% > %\vspace{-4mm}
% > \caption{Figure shows the the distances between neighboring neurons.}
% > \label{fig6}
% > \end{figure}
% > SOM or self-organizing map is Neural network that have a set of neurons connected to form a topological grid.
% 
% ^.
%
\begin{figure}[!h]
\centering
\includegraphics[width=260pt]{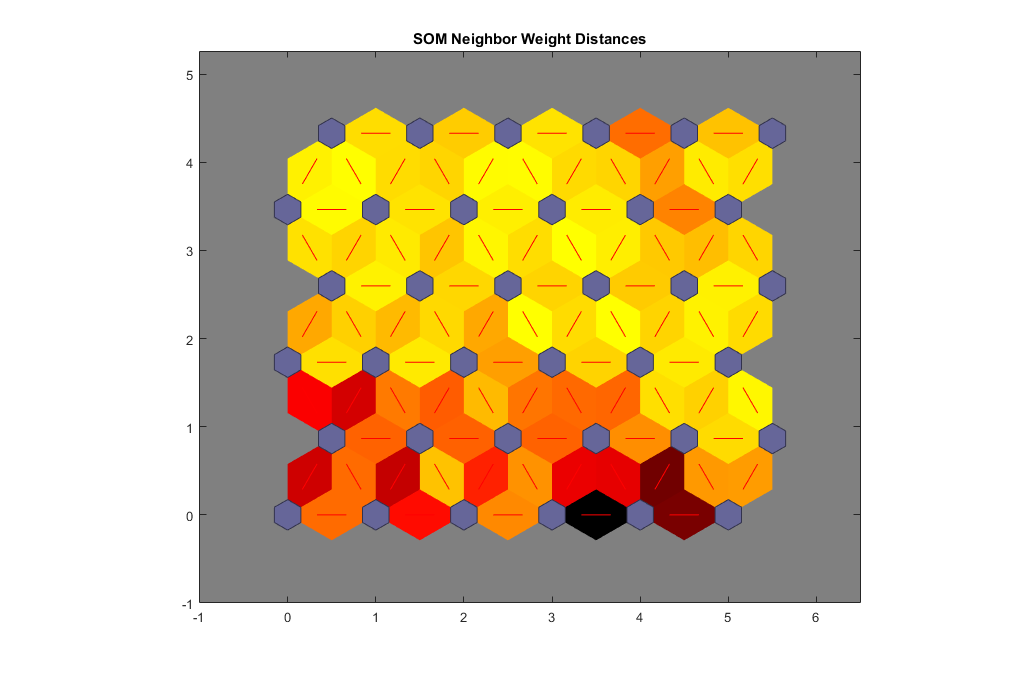}
%\vspace{-4mm}
\caption{Distances between neighboring neurons.}
\label{fig6}
\end{figure}
\begin{figure}[!h]
\centering
\includegraphics[width=250pt]{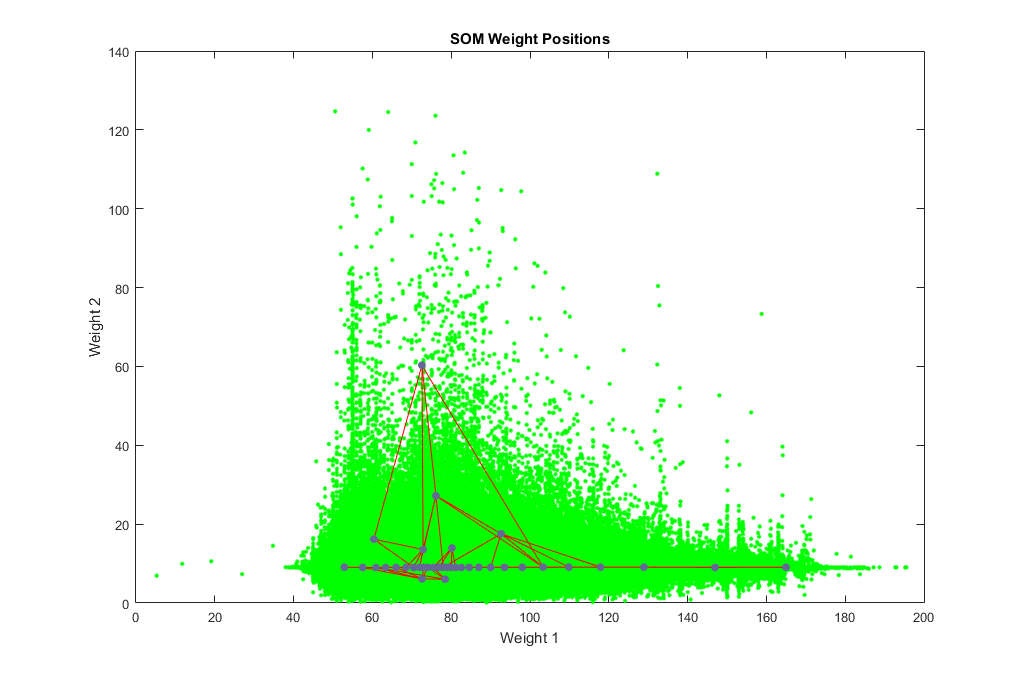}
% \includegraphics[width=1.0\columnwidth]{image7.png}
%\vspace{-4mm}
\caption{Data point locations and weight vectors.}
\label{fig7}
\end{figure}
SOM or self-organizing map is a neural network method with a set of neurons connected to form a topological grid. We got a topographical map of input patterns of the $AnEAR$ data with Acceleration and Heart rate as feature space as shown in Figure \ref{fig6}.
SOM was trained with a Neural Network of 36 Neurons. Each Neuron has several data-points in it, and thus acts as its own cluster. The SOM pattern is shown as a hexagonal grid. The blue hexagons in the figure represent the neurons, and the red lines connect neighboring neurons in the figures. The colors in the regions containing the red lines indicate the distances between neurons. Darker colors represent larger distances, and lighter colors represent smaller distances.  A group of Dark segments appears in the upper-right region, bounded by some lighter segments of yellow color. %Check the orientation of the figure, because I don't see that.
From the figure, this grouping might indicate that the network has clustered the data into three groups. The color difference indicates that data points in a particular region are farther apart. 

We had two parameters so we have seen two weights. The dataset consist of two features from ten subjects. Those features were Heart Rate and average Acceleration.
It is hard to see distinguishable clusters from the plot of SOM weight positions, but we can clearly see the distribution of the data set and the neurons used to learn to classify input vectors according to their grouping in the input space.
The blue dots or circles represent the neurons and green ones represent the data points.

Clustering analysis can reveal hidden patterns in the underlying data. We have shown three different types of clustering in this paper: \textit{k}-means, Gaussian Mixture Model clustering and Neural Network based SOMs. \textit{K}-means used 3 Clusters to group the $AnEAR$ data and GMM used 2 Clusters. While SOM did not use a distinct number of clusters, instead using 36 neurons, figure \ref{fig6} suggests the presence of 3 clusters in the data.
The smartwatch dataset is rich and can provide vital information about the person wearing the device.
\section{Conclusion}
%\vspace{0.2cm}
%
In this study, we leveraged multi-modal unlabeled smartwatch data from the $AnEAR$ framework.
We have shown the possibility of the correlation between different modalities could be helpful in predicting the user's activity or neurological state. We discussed results obtained through various clustering approaches such as K-means clustering, GMM clustering and SOMs for discovering  the latent structures in multi-modal smartwatch data. As a future work, it would be interesting to study the analysis of such a system on data obtained through patients.
\section*{Acknowledgement}
%\vspace{0.1cm}
%
We appreciate Alyssa Zisk who helped with proofreading of this manuscript. This research was supported by the National Institutes of Health (NIH) grant (no: R01MH108641).
\vspace{0.2cm}
%LOL 
                                  % on the last page of the 
%%%%%%%%%%%%%%%%%%%%%%%%%%%%%%%%%%%%%%%%%%%%%%%%%%%%%%%%%%%%%%%%%%%%%%%%%%%%%%%%
%
\bibliographystyle{IEEEtran}  
\bibliography{g}
\end{document}